# Towards trustworthy management of AIGC copyright: blockchain-enabled full lifecycle recording and multi-party auditing approach

Jiajia Jiang[1], Moting Su[2*], Aiqun Wu[3], Fengshu Li[4], Xiangli Xiao[5,6] and Yushu Zhang[5,6]

**Abstract**

With the escalating proliferation of artificial intelligence technologies, AI-generated content (AIGC) has progressively permeated across diverse domains. However, this explosive application has also sparked widespread public discussion about the copyright of AIGC. Existing copyright legal frameworks, originally designed around human creators, now face a paradigm shift. As human involvement in the generation of AIGC diminishes, where creative expression increasingly hinges on AI. This discrepancy has introduced multifaceted complexities and challenges in determining the copyright ownership of AIGC within established legal boundaries. Given this, meticulous recording and auditing of contributions from all parties in AIGC generation becomes imperative. Blockchain, with its decentralized storage, offers a robust technical foundation for AIGC copyright management. Yet existing blockchain-based solutions have clear limitations: most only focus on certifying final generated products, ignoring the management of critical intermediate data across the full lifecycle, thus failing to meet the needs of core scenarios like copyright confirmation and multi-party profit distribution. For this purpose, this paper introduces AIGC-Chain, a trustworthy AIGC copyright management system. It conducts a comprehensive recording of intermediate data generated across the full lifecycle of AIGC. Such data is deposited into a decentralized blockchain for secure multi-party auditing, thereby constructing a trustworthy management for AIGC copyright. In copyright dispute scenarios, auditors can retrieve critical proof from the blockchain, facilitating precise determination of the copyright ownership of AIGC products. Both theoretical and experimental analyses confirm that this scheme shows exceptional performance and security in AIGC copyright management.

*Correspondence:
Moting Su
sumoting@jxufe.edu.cn
[1] The College of Computer Science and Technology, Nanjing University of Aeronautics and Astronautics, Nanjing 210016, China
[2] The School of Business Administration, Jiangxi University of Finance and Economics, Nanchang 330013, China
[3] Shanghai Aerospace Information Technology Research Institute, Shanghai 201109, China
[4] The School of Business, Anhui University, Hefei 230601, China
[5] The School of Computing and Artificial Intelligence, Jiangxi University of Finance and Economics, Nanchang 330013, China
[6] The Jiangxi Provincial Key Laboratory of Multimedia Intelligent Processing, Nanchang 330044, China

## Introduction

The widespread adoption of technologies such as AI painting and intelligent chat software like ChatGPT[1] has drawn increased attention to Artificial Intelligence-Generated Content (AIGC) and has led to its broad application across various domains (Luo et al. 2022; Sun et al. 2022; Kandlhofer et al. 2016). AIGC leverages artificial intelligence techniques, including Generative Adversarial Networks (GAN) (Krichen 2023) and diffusion models, to train on existing datasets and generate new content based on specific input prompts. In 2018, an artwork created by artificial intelligence was sold at Christie's auction house for $432,500,[2] marking the first-ever sale of an AIGC product worldwide. The price achieved underscored the commercial and cultural significance of AIGC. However, this sparked a significant debate on AIGC copyright ownership, involving the distribution of rights among model providers, users, and other possible creative contributors of AIGC (He 2019).

[1] https://openai.com/chatgpt/

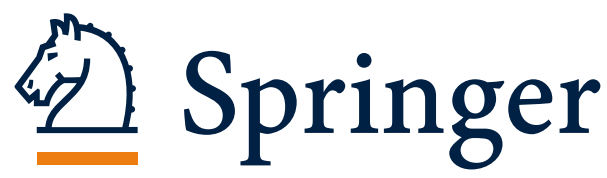

Copyright refers to the legal rights of creators to their literary, artistic, and scientific works, intending to protect the creative intellectual output of creators from unauthorized use by others, and encouraging innovation and cultural development (Liang 2015). The substitution of human creative activities by AI has posed new challenges to the existing copyright regime. Traditional copyright law incentivizes the generation and dissemination of works by conferring statutory exclusive rights on right holders. However, AIGC constitutes a new category of *orphan works*. Specifically, the generation process of AIGC differs significantly from traditional human creation: unlike products directly produced by human creators based on their own inspiration, conception, and creative skills, AIGC is generated by AI programs. This unique mode of creation has undoubtedly raised profound questions regarding copyright ownership: whether the copyright should belong to the AI itself, model providers, users, or other relevant parties–these issues urgently require deep exploration and clear delineation (Lu 2025). In contrast, the existing legal framework is predicated on the assumption of natural person creation, which presents challenges to the determination of authorship, the requirement of originality, and the design of protection pathways for AIGC. Meanwhile, copyright law generally mandates that works possess a certain degree of originality. Yet, whether AIGC meets this criterion and how to assess the originality of AI-generated content remain ill-defined issues (Gaffar and Albarashdi 2025), frequently placing AIGC in quandaries related to copyright infringement.

In November 2023, Beijing Internet Court concluded a copyright infringement lawsuit,[3] in which the plaintiff submitted a comprehensive record of prompts and parameters employed in the content creation process, effectively establishing that the generated images reflected human creative intellectual input. Consequently, the court ruled in favor of the plaintiff, granting them the ownership of copyright over the image. This case illustrates that it is necessary to accurately record the key parameters and operational steps involved in the generation of AIGC products to ensure the copyright ownership of the generated AIGC product. Such records can serve as critical evidence to substantiate the intellectual and knowledge-intensive efforts embedded in the generation process, thereby supporting the legitimacy of copyright claims.

However, traditional data storage methods, with their centralized architecture, are vulnerable to attacks and tampering (Jiang et al. 2022; Wei et al. 2025). It allows each participant in the process of AIGC to potentially submit altered or fabricated proof that favors their interests during copyright attribution, affecting the determination of copyright ownership for AIGC (Zhang et al. 2023). To address this flaw, there is an urgent need for a more secure and transparent recording system that ensures the integrity of the proof. In this context, the application of blockchain appears to be crucial. Blockchain, with its unique decentralized structure and advanced encryption algorithms, can provide AIGC with a highly robust, decentralized, and manageable end-to-end copyright management platform (Jiang et al. 2025; Ramzan et al. 2023). Storing the information of all key steps in the full lifecycle of AIGC in the blockchain ensures the reliability and security of AIGC and effectively resists any unauthorized tampering behaviors (Franke et al. 2024). Moreover, the timestamp of blockchain assigns an exact temporal verification to each record, enhancing the integrity and credibility of the proof while also ensuring the immutability of the transaction timeline, thus providing strong technical support for the management and legal defense of AIGC copyrights (Baliker et al. 2024).

So far, significant advancements have been achieved in the application of blockchain for copyright protection (Liang et al. 2021; Chen et al. 2023; Ansori et al. 2023; Wang et al. 2022; Natgunanathan et al. 2022; Meng et al. 2018), particularly in enhancing the security and traceability of copyright information storage. However, the majority of these applications are tailored to scenarios where the attribution of copyright is unambiguous. The copyright of each AIGC product involves the contributions of multiple participants throughout various stages of content generation, which are often difficult to delineate and quantify. Evidently, the existing blockchain-enabled copyright protection strategies have not adequately addressed this complexity. Currently, research efforts are underway to apply blockchain to the full-lifecycle management of AIGC copyright (Liu et al. 2024a, b). Although they have provided valuable insights into addressing issues related to AIGC, there remains a lack of methods to manage AIGC copyright from a global perspective, which makes it difficult to solve problems such as unclear ownership of AIGC and the difficulty in defining rights and obligations.

[2] https://edition.cnn.com/style/article/obvious-ai-art-christies-auction-smart-creativity/index.html.

[3] https://patentlyo.com/media/2023/12/Li-v-Liu-Beijing-Internet-Court-20231127-with-English-Translation.pdf.

In this paper, we facilitate trustworthy AIGC copyright management by meticulously recording intermediate data generated throughout the full lifecycle of AIGC and leveraging blockchain to achieve decentralized multi-party auditing. As illustrated in Fig. 1, we pre-recorded the intermediate data generated at each stage of the AIGC lifecycle on the AIGC-Chain. In the instance of copyright disputes, we draw upon the data from the decentralized ledger to provide proof, thus offering a solid foundation and support for resolving copyright conflicts. Specifically, we maintain a world state with eight attributes for each AIGC product, ensuring that intermediate data of the generation process is securely stored on the blockchain in an immutable manner. This methodology establishes a robust framework ensuring traceability, verifiability, and integrity of AIGC products, effectively overcoming the constraints inherent in conventional content generation paradigms. Moving beyond both the anthropocentric authorship model and reductive instrumental approaches, it creates a reliable infrastructure for synergistic human-AI co-creation. Our main contributions can be summarized as follows:

- We introduce AIGC-Chain, a trustworthy copyright management system for the full lifecycle of AIGC, meticulously designing five transaction phases required for documentation and details of their state transition on the blockchain, aiming to achieve holistic management and complete recording of the status of AIGC.
- We propose a decentralized multi-party auditing approach with blockchain, which enables automated copyright proof collection. It is augmented by IPFS

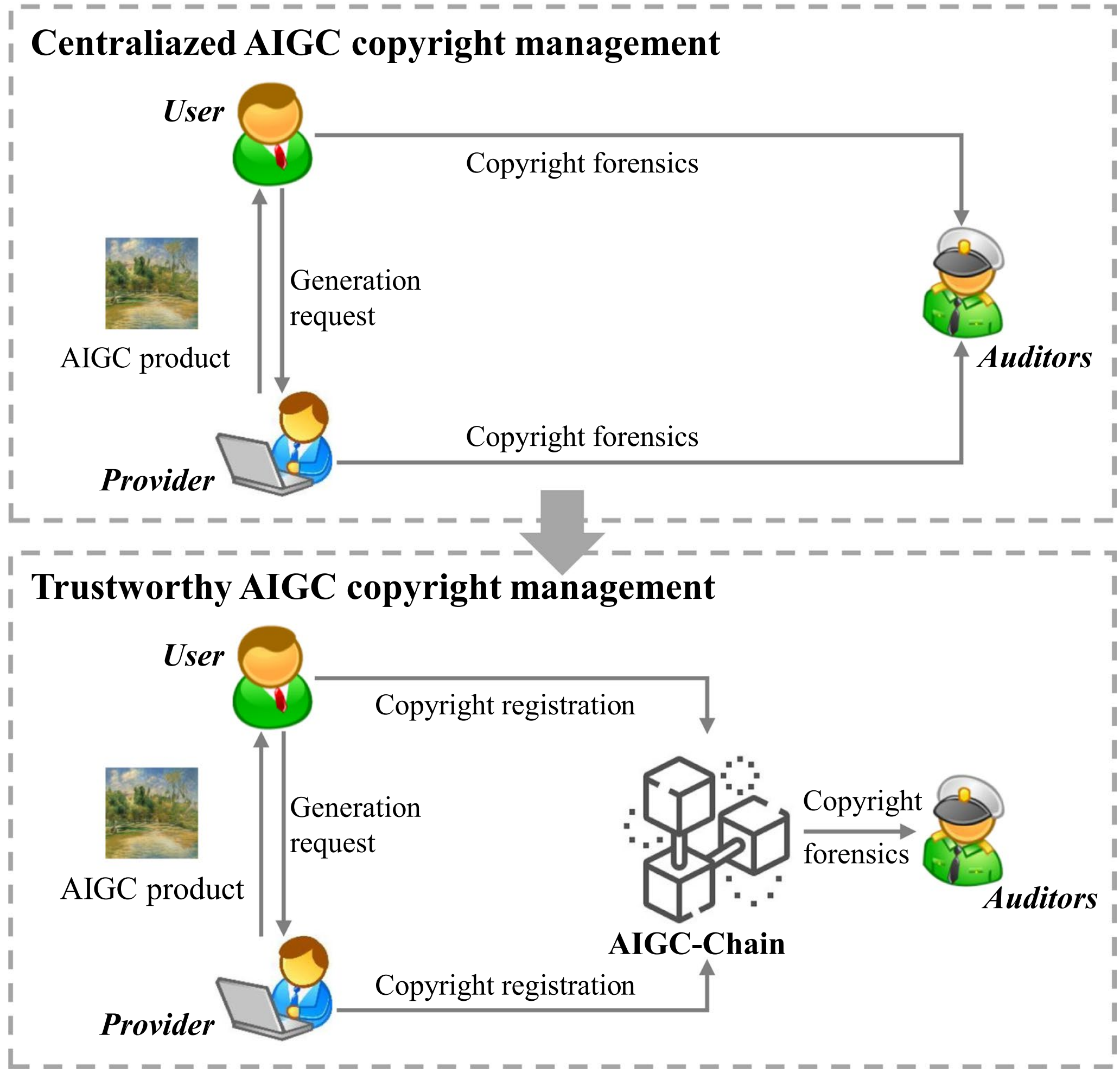

**Fig. 1** Graphical abstract of AIGC-chain

technology to safeguard the security and reliable storage of AIGC copyright.

- We develop *CTrace*, a copyright tracing algorithm based on bloom filters. It improves the efficiency and accuracy of transaction query and verification within the blockchain and provides strong evidence for resolving copyright disputes, enabling trustworthy management of AIGC copyright.

The remainder of this paper is organized as follows. "Relatedwork" section reviews existing works that are related to ours. "Preliminaries" section introduces the preliminary used in the proposed system. "Models" section defines the system model, threat model, and design goals of AIGC-Chain. "The proposed system" presents the detailed construction of AIGC-Chain. "Theoretically evaluation" section assesses the achievement of design goals. "Performance evaluation" section evaluates the performance of AIGC-Chain. "Conclusion" section concludes this paper and points out the future research directions.

## Related work

This section first systematically analyzes core risks in AIGC copyright management, then sorts out existing research from three dimensions: legal revision, technical protection, and full-lifecycle responsibility division, aiming to provide a theoretical reference for building a standardized and efficient AIGC copyright management system.

### Core risks in AIGC copyright management

AIGC copyright risks are multi-dimensional, violating original legitimate rights of generators and disrupting the digital creative market order (Zhang et al. 2023). These risks manifest in three key aspects:

*Training data copyright infringement.* AIGC models depend on massive copyrighted data for training, yet most developers lack explicit authorization from copyright owners. This unauthorized use potentially infringes on reproduction and information network dissemination rights (Mochram et al. 2022). For instance, image-generating tools often use copyrighted works without permission, while large language models crawl copyrighted literary works and academic papers. Moreover, opaque training data sources make it hard for rights owners to trace unauthorized use and provide infringement evidence, leading to a difficult situation of presenting evidence in copyright protection.

*Attribution ambiguity of AIGC outputs.* The traditional copyright system clearly defines eligible subjects (natural persons, legal persons, etc.) and acquisition conditions (originality, reproducibility). However, AIGC breaks the paradigm of human creation, causing attribution ambiguity. On one hand, AIGC lacks legal personality and cannot be a copyright subject under current laws. On the other hand, copyright ownership among stakeholders remains unclear. This ambiguity triggers frequent disputes and hinders the market value realization of AIGC (He 2019).

*Difficulties in supervision and enforcement.* The high efficiency, massiveness, and diversity of AIGC outputs create three major supervision challenges. First, its fast generation speed renders traditional manual supervision ineffective. Second, blurred boundaries between AIGC and originals complicate infringement identification, as subjective judgment struggles to determine substantial similarity. Third, cross-border dissemination of AIGC leads to complex enforcement issues–differences in national copyright laws and standards result in jurisdiction conflicts, cross-border evidence collection difficulties, and low enforcement efficiency, severely undermining copyright protection effectiveness (Liu et al. 2024a).

### Existing research on AIGC copyright management

In response to the aforementioned risks, some research has been conducted, forming a series of findings centered on the optimization of legal systems, the formulation of general techniques for AIGC management, and the construction of full-lifecycle management mechanisms (Ye et al. 2025). This section consolidates and summarizes existing research from the following three perspectives:

#### *Legal revision*

AIGC is produced by AI programs rather than human creators, its copyright has become a subject of intense debate. Some scholars contend that the copyright of AIGC should be assigned to the developers or proprietors of the AI programs, whereas others argue that AIGC should be treated as a public resource accessible for unrestricted use by anyone. From an academic research perspective, relevant studies have explored this issue from multiple dimensions. Gaffar and Albarashdi (2025) discussed the copyright of AIGC from a legal standpoint, attempting to clarify the definition of copyright applicable to AIGC and the specific methods for determining its ownership. Mei (2025) systematically sorted out the focal disputes regarding AIGC copyright from a legal perspective. Aiming at the dilemma of copyright qualification specific to generative content, they proposed a path for improving the copyright system, which provides a legal analytical framework for defining the copyright boundary of AIGC. In contrast, Ye et al. (2025) held that the input of intellectual labor by users is the key determinant of AIGC copyright ownership. Taking various participants in the AIGC generation as the analytical dimension, they analyzed the types of copyright risks faced by different subjects, and constructed a collaborative governance scheme from the aspects of legal regulation, technical support, and responsibility division. In terms of practical legal practice, several countries, including the United Kingdom, Ireland, New Zealand, and India, have proactively initiated legal reforms to more effectively address copyright challenges related to works that incorporate computer-generated elements and those created with the assistance of artificial intelligence (Wan and Lu 2021). The revision of these regulations provides a solid legal framework for the management of AIGC, while facilitating the exploration of technical approaches to its targeted administration.

#### *General techniques in AIGC management*

Technical protection serves as a crucial method to compensate for the limitations of legal regulation and enhance the efficiency of AIGC copyright management. Existing research mainly focuses on the application of technologies such as blockchain and digital watermarking. Scholars propose leveraging the immutability, traceability, and transparency of blockchain to establish a decentralized AIGC copyright management platform (Zhang et al. 2023; Lockl et al. 2020; Xiao et al. 2023). Specifically, once an AIGC product is generated, relevant metadata is immediately recorded on the blockchain to form tamper-proof copyright certificate, which provides a reliable basis for clarifying copyright ownership and identifying infringement behaviors. Iyengar et al. (2023) performed a thorough examination to evaluate the advantages and difficulties of applying blockchain in supply chain management. Xiao et al. (2023) proposed a fair payment scheme based on blockchain, which has been proven to ensure the secure transaction of copyrighted content effectively. Chen et al. (2023) introduced a blockchain-based proactive defense strategy that simplifies the copyright transfer process but also realizes the pre-warning and punishment of infringement behaviors.

Digital watermarking has emerged as another key research direction. Researchers are dedicated to developing digital watermarks tailored for AIGC scenarios by embedding invisible watermarks into training data or AIGC products. Watermarks can still be accurately extracted even after the products undergo editing or transformation, realizing the full-process tracing of AIGC products from generation to dissemination. Ren et al. (2024) systematically analyzed the application status and potential security risks of digital watermarking technology in the field of AIGC from a multimodal perspective (Yang et al. 2022, 2024; Fernandez et al. 2023; Liu et al. 2023; Zhang et al. 2023), and further emphasized the core challenges currently faced by this field.

However, the original intention of these solutions primarily focuses on realizing the traceability of AIGC products, but fails to address the issue of ambiguous copyright ownership caused by the varying contribution degrees of participants throughout the full lifecycle of AIGC.

#### *Full lifecycle management for AIGC*

The AIGC generation process involves multiple participants, including training data providers, model developers, AIGC content users, and distribution platforms. Currently, one of the core root causes of chaos in AIGC copyright management is the ambiguous division of responsibilities among these participants. Therefore, constructing a full-lifecycle AIGC copyright management system has become a core research direction for clarifying the copyright boundaries and responsibility scopes of each participant. Chen et al. (2023) proposed a blockchain-enabled approach for AIGC ownership verification. By establishing a reliable AIGC ownership verification system, it provides technical support for the secure training, deployment, and circulation of AIGC. Liu et al. (2024a) addressed the core challenges in the AIGC edge network environment by designing a blockchain-based AIGC management scheme, which achieves key functions such as content tamper-proofing, anti-plagiarism, and secure transactions. In addition, they established a blockchain-driven AIGC model evaluation mechanism (Liu et al. 2024), which mitigates AIGC security risks arising from irregular model generation behaviors through the design of differentiated incentive policies. Yang et al. (2024), on the other hand, proposed a blockchain-based AIGC regulatory framework that focuses on safeguarding identity information security for owners of AIGC products. While the aforementioned studies have offered valuable insights for addressing AIGC-related security issues, there is still a lack of methods for managing the full lifecycle of AIGC from a global perspective, making it difficult to solve the problems of ambiguous AIGC copyright ownership and the difficult delimitation of rights and obligations.

## Preliminaries

In this section, we begin by exploring the blockchain technology utilized within AIGC-Chain. Then, we introduce the Indistinguishable Bloom Filter, which is used in our tracing method for transactions.

### Consortium blockchain

A consortium chain is a blockchain network controlled by a limited number of organizations, known as consortium members. These members agree on the rules and protocols governing the network and work together to validate transactions and maintain the integrity of the ledger (Kim et al. 2022). Unlike public blockchains, where anyone can join and participate, consortium chains require explicit invitation and membership.

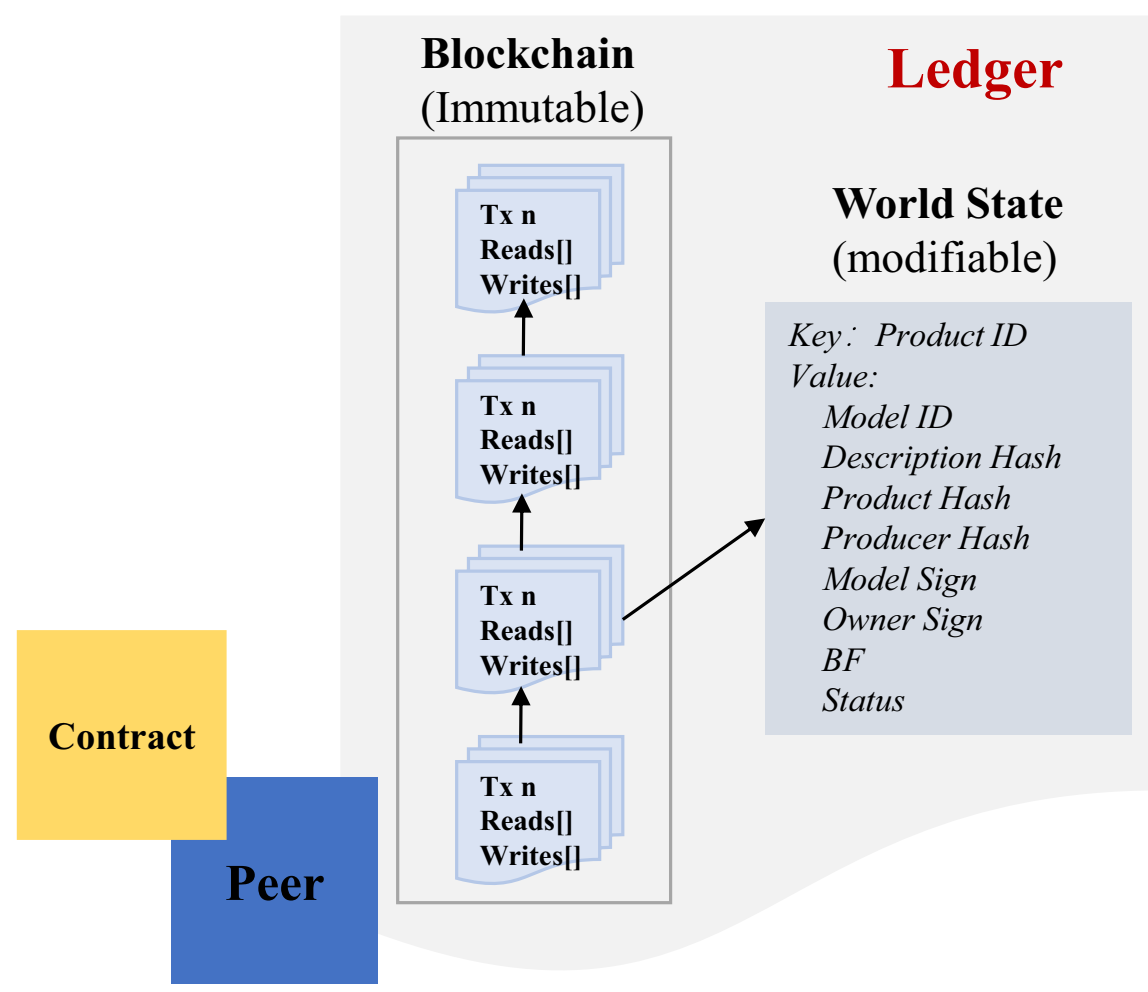

**Fig. 2** The structure of consortium blockchain network

As illustrated in Fig. 2, the blockchain network is composed of orderer nodes and peer nodes. The orderer node is responsible for sorting transactions and distributing them to all peer nodes on relevant channels. Peer nodes, in turn, host both the ledger and smart contracts. The ledger maintains a systematically organized and tamper-proof record of all transactions. Additionally, the ledger incorporates a state database that tracks the current status, known as the world state. Every transaction generates a collection of key-value pairs that are utilized to update the world state.

The client sends a transaction proposal to one or more orderer nodes in the network. Orderer nodes pre-sort the transaction and validate its legitimacy. After pre-sorting, orderer nodes distribute the transaction to all relevant peer nodes on the channel. Upon receiving the transaction, peer nodes execute query or modification operations specified in the transaction and send the results back to orderer nodes. Orderer nodes collect the transaction results from all peer nodes and confirm the transaction. The transaction is then added to the blockchain of the channel and synchronized among all peer nodes in the network. The client application can query the transaction result to confirm that the transaction has been executed correctly.

#### Indistinguishable bloom filter

Indistinguishable Bloom filter (IBF) (Liang et al. 2023) obfuscates data by simulating future unknown queries. As an extension of the Bloom Filter (BFs) (Tarkoma et al. 2012), IBF ensures efficient and accurate data retrieval while safeguarding data privacy. As illustrated in Fig. 3, IBF is constructed from an $m$-bit Twin Bloom Filter (Li and Liu 2017). Within each twin, one cell is designated as chosen and the other as unchosen, with the unchosen cell potentially holding either 0 or 1. IBF encompasses three primary algorithms: *IBF.Setup*, *IBF.Initi*, *IBF.Insert*, and *IBF.Check*.

- *IBF.Setup*$(1^{\eta}, k, m) \rightarrow \{H_{key}, H\}$: Construct $k$ pseudo-random hash functions $H_{key} = \{h_1, h_2, \cdots, h_{k+1}\}$ using the keyed-hash message authentication code (HMAC) and a hash function $H(x) = \mathbb{H}\%2$, where $\eta$ is the security parameter, $k$ is the number of hash functions, and $\mathbb{H}$ is a pseudorandom hash function.
- *IBF.Initi*$(m, H_{key}, H, \gamma \rightarrow IBF)$ Calculate $H(h_{k+1}(j) \oplus \gamma)$ to determine which cell in each twin is chosen, where $\gamma$ is a random number and $j \in \{0, m-1\}$.
- *IBF.Insert*$(IBF, v, H_{key}, H, \gamma) \rightarrow IBF$: For each value $v \in V$, set $IBF[(h_i(v))][H(h_{k+1}(h_i(v)) \oplus \gamma] = 1$ and $IBF[(h_i(v))][1 - H(h_{k+1}(h_i(v)) \oplus \gamma] = 0$, where $i \in [1, k]$.
- *IBF.Check*$(IBF, v^{'}, \gamma) \rightarrow 0/1$: Check whether $IBF\left[(h_i(v^{'}))\right]\left[H(h_{k+1}(h_i(v^{'})) \oplus \gamma\right] = 1$ for the queried value $v^{'}$. If the values of $k$ chosen cells are 1, $v^{'}$ is included in IBF; otherwise, not included.

The correctness of IBF is established if, for all parameters $\eta$, $m$, $k$, $\gamma$, and any value $v$, following the execution of *IBF.Setup*, *IBF.Init*, *IBF.Insert*, and *IBF.Check*, the function *IBF.Check* returns a value of 1 when $v' = v$, otherwise

$$
\begin{aligned}
&Pr[IBF.Check(IBF, v^{'}, \gamma)] = 0] \\
&\quad = 1 - \mathrm{negl}_{\mathrm{IBF}}(v, k, |V|),
\end{aligned}
$$

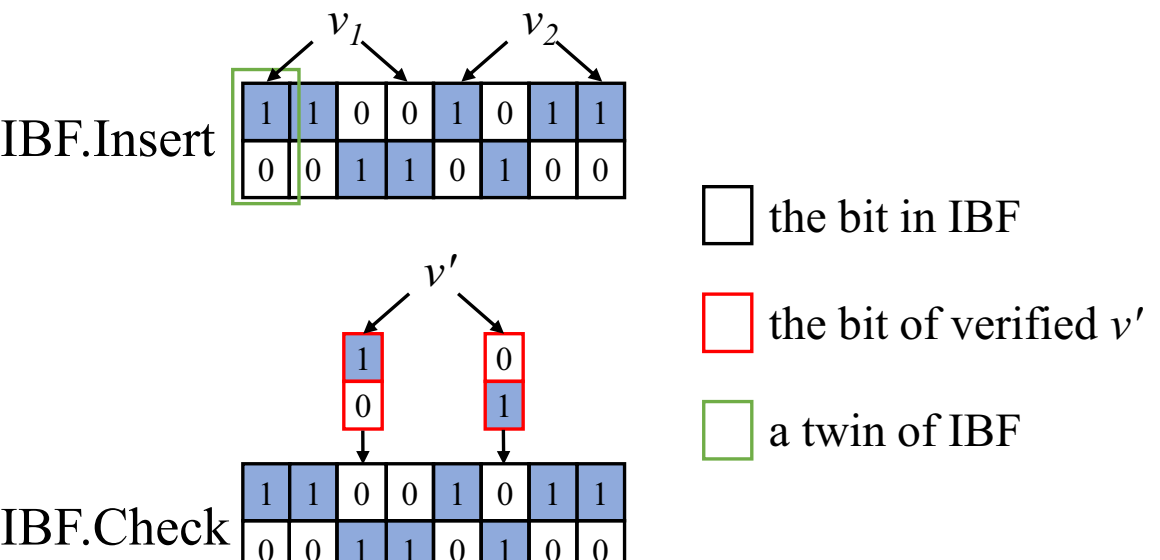

**Fig. 3** The technical principle of indistinguishable bloom filter

where the false positive rate of IBF is identical to that of traditional BF, i.e., $\text{neglIBF}(m, k, |V|) = \text{neglBF}(m, k, |V|) \approx (1 - (1 - \frac{1}{m})^{k|V|})^k \approx (1 - e^{\frac{-k|V|}{m}})^k$.

## Models

This section starts by introducing the roles in the AIGC-Chain and provides a brief overview of the process design of the proposed system. Then, it elaborates on the threat model and design goals.

### System model

Figure 4 depicts four key roles within AIGC-Chain: *User*, *Provider*, *Consumer*, and *Auditor*. These roles interact with AIGC-Chain by submitting transactions to record copyright information on the blockchain, facilitating secure management of AIGC copyright details. Here are the detailed descriptions of the four roles:

- *User*: *User* is the initiator of AIGC generation, existing as a light node on AIGC-Chain. It selects an appropriate model and inputs a prompt to trigger the generation process. *User* is the initial owner of the AIGC product, with the right to use, share, and transfer it.
- *Provider*: *Provider* owns the trained AIGC model and exists as a full node. It offers AIGC generation services to *User*s and charges a fee for these services. An important responsibility of *Provider* is to promptly record the usage and generation activities of models on AIGC-Chain.
- *Consumer*: *Consumer* is the entity who plans to purchase or utilize the AIGC product, existing as a light node. It can retrieve information about the AIGC product from the chain and propose a transaction to modify the ownership of the AIGC product.
- *Auditor*: *Auditor* represents copyright certification institutions, courts, and other authoritative bodies, existing as a full node. It gathers proof from AIGC-Chain to adjudicate copyright disputes related to AIGC and ensure fair and just outcomes.

### Threat model

The threats considered in our scheme are as follows, which come from four entities, i.e., *User*, *Provider*, *Consumer*.

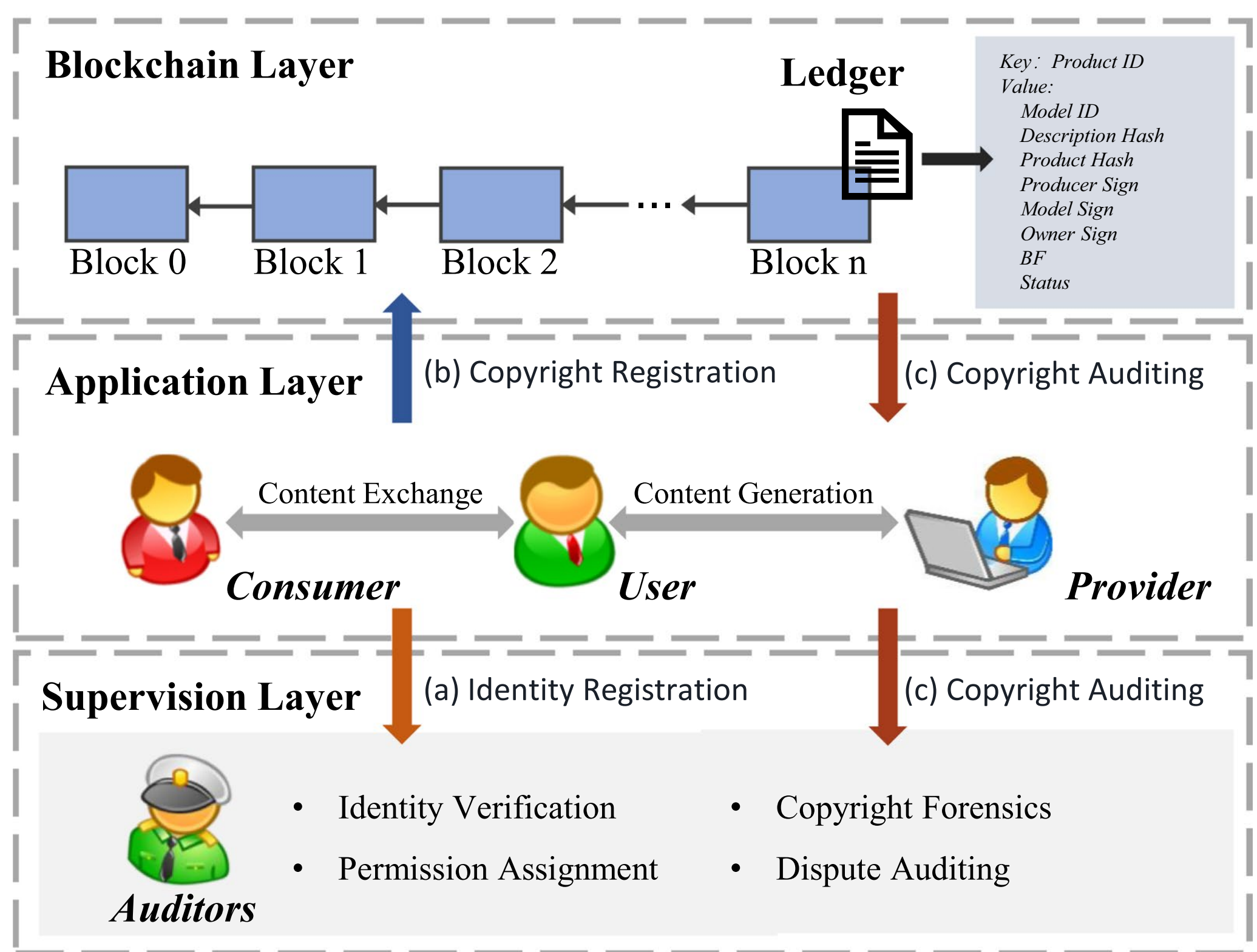

**Fig. 4** System model

- *User*, *Provider*, and *Consumer* may intentionally submit false or misleading proof to claim copyright ownership over the AIGC product.
- *User* and *Provider* might furnish the *Auditor* with erroneous or fabricated transaction ID, which could lead to incorrect copyright assessments for AIGC products, thereby misleading the *Auditor* and preventing the tracing of the accurate copyright information related to the AIGC products.
- During the copyright audit of the AIGC product, *Auditor*s may collude with others, potentially leading to erroneous or unfair audit outcomes.

The threats beyond the scope mentioned above fall outside the scope of our primary treatment. These include activities such as eavesdropping on communication channels and attacks on blockchain node servers. These issues can be addressed by incorporating general solutions into our framework.

### Design goals

In response to the provided system model and threat model, the intended design goals are summarized as follows:

- *Trustworthy*: Throughout each phase of generation, the intermediate data outputs are meticulously and accurately uploaded to the AIGC-Chain. By harnessing the inherent immutability of blockchain, we guarantee the trustworthiness of the copyright proof retrieved, facilitating the management of copyright ownership.
- *Traceability*: All data generated throughout the full lifecycle of an AIGC product can be permanently recorded and cryptographically verifiable, eliminating risks of tampering or unauthorized modifications to copyright-related information.
- *Fairness*: Authoritative regulatory entities such as copyright certification institutions and courts are introduced as *Auditor* in AIGC-Chain. When copyright disputes arise, these nodes can retrieve proof from AIGC-Chain to ensure unbiased and fair judgments regarding the copyright of AIGC products.

## The proposed system

In this section, we delve into the specifics of the proposed system. We begin by providing a comprehensive overview. Subsequently, we introduce a novel approach for tracing data within the blockchain. We conclude by elaborating on the operational workflow and the corresponding state transitions within the proposed system. For improved readability, a concise compilation of the key notations used in this section, including their definitions, is provided in Table 1.

**Table 1** Notations and descriptions

| Notations | Descriptions |
|---|---|
| $P$ | The AIGC product |
| $P'$ | The AIGC product to be verified |
| *Args* | The prompt for generating AIGC product |
| *Product ID* | The key for an AIGC product in the world state |
| *Model ID* | The unique identifier for the used model |
| *Description hash* | The hash of the input prompt *Args* |
| *Product hash* | The hash of the AIGC product $P$ |
| *User sign* | The signature of *User* |
| *Model sign* | The signature of *Provider* |
| *Owner sign* | The signature of the owner of the AIGC product |
| *BF* | The index of all transactions of the AIGC product |
| *Status* | The current status of the AIGC product |

### Overview

By storing data generated throughout the AIGC lifecycle in a complete and unchangeable form on a distributed blockchain, we can ensure the trustworthy and traceability of the sources of AIGC products. This also guarantees that in the event of copyright disputes, data stored on AIGC-Chain can be used to determine the ownership of AIGC products, thereby protecting the security of AIGC copyright. As depicted in Fig. 4, AIGC-Chain comprises 5 core procedures: *Registration*, *Content Generation*, *Data Uploading*, *Copyright Trading*, and *Copyright Management*.

AIGC-Chain is deployed on a consortium blockchain network, which maintains its data in a decentralized blockchain ledger composed of blocks and the world state. Each block contains records of all preceding transactions that have been successfully executed, while the world state preserves the latest data state in the form of key-value pairs. The adoption of consortium blockchain embodies a dynamic balance between technical feasibility and scenario suitability in AIGC copyright management. While public blockchains ensure proof traceability via full-network data transparency, they risk leaking core digital assets like user prompts and model parameters. Private blockchains, by contrast, lack a decentralized trust framework and thus fail to provide multi-party recognized third-party credibility in copyright disputes. Conversely, the core attributes of consortium blockchain–authorized access and distributed consensus–precisely address the adaptability flaws of public and private blockchains in AIGC copyright management scenarios. Its refined permission control system enables hierarchical, trusted collaboration among copyright participants, while the combined deployment of distributed consensus mechanisms and authoritative audit nodes can technically ensure the impartiality and compliance of copyright ownership verification and infringement evidence collection. These features render the consortium blockchain fully suited to the full-lifecycle demands of AIGC copyright management.

All participants in AIGC-chain are required to complete identity verification and qualification certification via a certificate authority (CA). The Proof of Authority (PoA) consensus mechanism abandons the computational power-intensive design of Proof of Work (PoW) and dispenses with the token-staking requirement of Proof of Stake (PoS) for maintaining consensus stability, thus cutting operational costs and lowering participation barriers. Moreover, the deterministic verification of authorized nodes markedly elevates consensus efficiency, which fully meets the high-frequency transaction demands of AIGC copyright management scenarios. This mechanism ensures full traceability in copyright management, mitigates the regulatory gaps arising from the open participation model of public chain consensus mechanisms, and enhances the credibility of AIGC copyright management and dispute resolution. Hence, PoA stands as the optimal solution for enabling AIGC-chain functionalities.

Within AIGC-Chain, only the key metadata supporting copyright verification, traceability, and transactions is stored on-chain to avoid the occupation of massive block space by raw data. High-frequency access data (e.g., product status, ownership signatures) is retained in the world state, whereas low-frequency access data (e.g., historical transaction indexes) is archived in block bodies. As illustrated in Fig. 4, a world state of 8 items is preserved in the ledger of AIGC-Chain for each AIGC product. This record is dynamically updated throughout the full lifecycle of the AIGC product, featuring a small single-record size and high structuralization, which enables efficient writing and querying to rapidly respond to various copyright operation requirements. Among them, *Model ID* serves as a unique identifier for the training models employed. Given that participants of the blockchain have full access to all data on the chain, we employ a strategy to protect the privacy of AIGC products. Instead of directly storing the original content of AIGC products, it generates descriptive hashes based on input parameters and updates the world state with these hashes (Heo et al. 2024). As a supplementary decentralized storage solution, the InterPlanetary File System (IPFS) is dedicated to storing large-capacity data that does not require real-time verification but demands long-term retention on-chain. This alleviates the storage pressure of the blockchain and ensures data traceability through its content addressing mechanism, with storage costs far lower than those of native blockchain storage. Additionally, *BF* is employed to index all transaction information related to the AIGC product on the blockchain, facilitating the authenticity management of copyright. Details of *BF* will be discussed in "Efficient tracing method based on indistinguishable bloom filter".

### Efficient tracing method based on indistinguishable bloom filter

As illustrated in Fig. 5, the generation of each AIGC product is inherently associated with at least three categories of on-chain transactions. In the event of a copyright dispute, *Auditor*s could access detailed copyright proof through *Product ID* and require all participants to submit copyright proofs of their involvement in the generation process, enabling fine-grained copyright traceability. It should be noted that while the input parameters *Args* in transactions can validate the authenticity of transactions, the transparent nature of blockchain allows potential attackers to forge transactions with identical *Args* in attempts to manipulate copyright attribution of AIGC products. Chen et al. (2023) proposed a singly-linked list-based fast query method, which uses local hash indexes to store copyright information hashes. It enables copyright verification without parsing full blockchain transactions, but its centralized storage requires maintaining offline index replicas, increasing local storage costs, and introducing single-point failure risks. Li et al. (2025) proposed a lightweight verification scheme based on Root Merkle Tree and Bloom Filter Merkle Tree, addressing traditional blockchain index issues, such as poor adaptability, high storage, and low efficiency. However, it relies on multi-round iterative training for convergence, causing initial performance fluctuations and high storage costs, and cannot support AIGC multi-participant contribution tracing. Hassanzadeh-Nazarabadi et al. (2021) designed a DHT-based index architecture, mitigating traditional index pain points, including high storage complexity, heavy burden, and slow new node onboarding. Yet, it suffers from linear storage overhead growth with increasing data scale, low throughput, and an inability to quantitatively trace in an AIGC scenario. Rasheed et al. (2022) developed a two-layer blockchain index based on zero-knowledge proof to solve privacy leakage, data manipulation, and identity forgery for IoT devices. Restricted by the computational complexity of ZKP, it has high computational and practical application overhead.

To address this vulnerability, we propose *CTtrace*, an efficient, copyright-traceable verification mechanism based on indistinguishable Bloom filters. It constructs an $m$-bit twin array *BF* for each blockchain-recorded

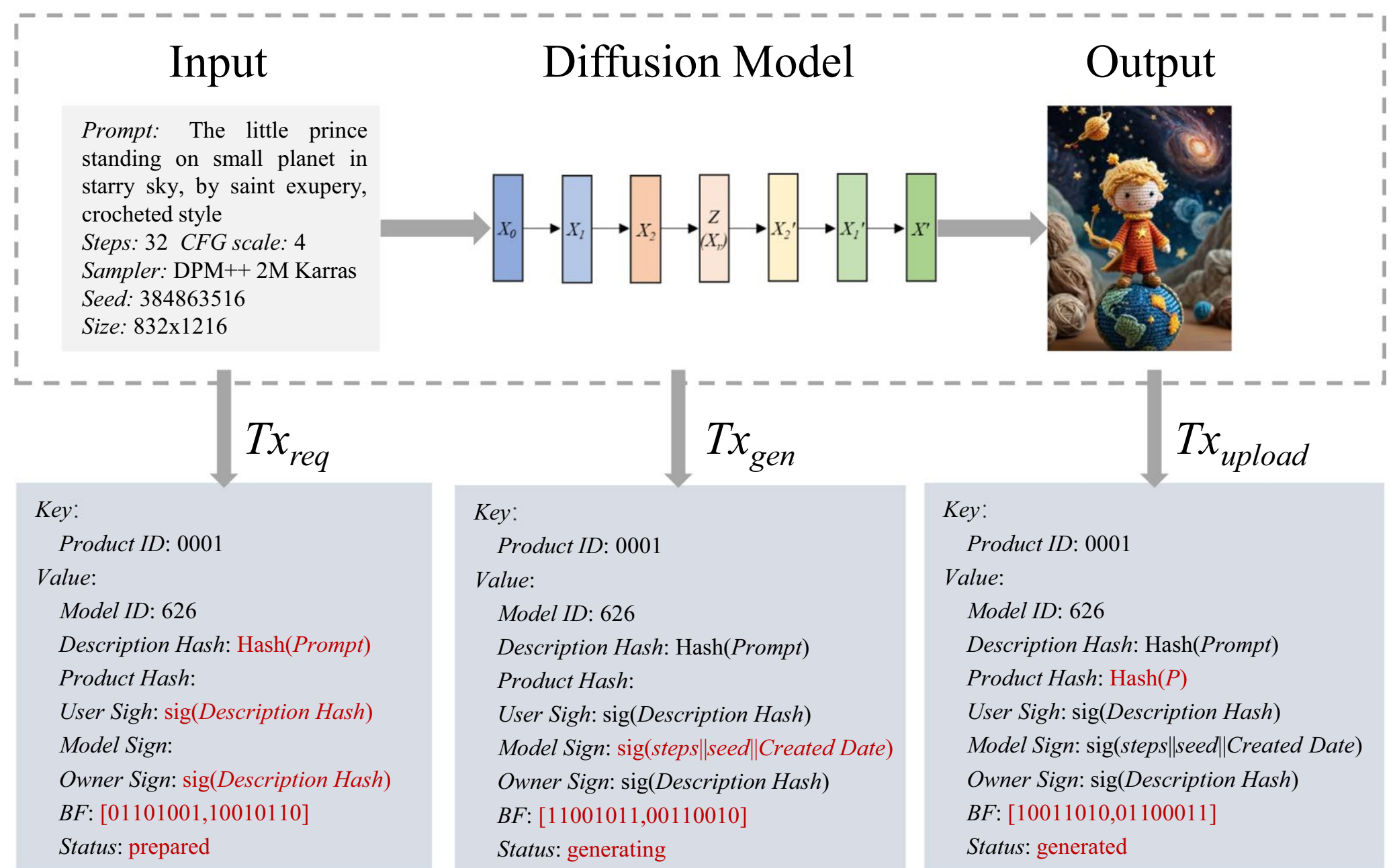

**Fig. 5** Lifecycle management on the blockchain

AIGC product with all associated transactions mapped. By implementing hash mapping of input parameter *Args*, timestamp *CreateTime*, and random value *Nonce*, *CTtrace* ensures that even if attackers duplicate *Args*, they cannot reconstruct legitimate transaction records with identical timestamps and nonces. It achieves constant-level storage overhead and ensures robust privacy protection through data obfuscation. More crucially, it exhibits excellent compatibility with AIGC copyright frameworks. It fully supports the traceability of intermediate processes across the full lifecycle and the quantification of multi-participant contributions, enabling robust binding between AIGC and its on-chain transactions, effectively precluding false positive results from being admitted as valid legal evidence in AIGC copyright dispute arbitration.

Specifically, as depicted in Algorithm 1, *CTrace* comprises three stages: *Setup*, *Insert*, and *Check*. The smart contract automatically executes these stages upon triggering transactions for copyright storage and querying.

1) *Setup*: Upon the receipt of an AIGC generation request from *User*, a dedicated array *BF* and a set of hash functions $\{H_{key}, H\}$ are generated for product *P*. These elements are subsequently embedded within the transaction and recorded in the world state on the blockchain. Access to this information is granted exclusively to *Provider*, *Consumers*, and *Auditors* affiliated with product *P*.
2) *Insert*: When a transaction is initiated to update the status of product *P*, its *BF* in the world state will be revised to:

$$BF[(h_i(v))][H(h_{k+1}(h_i(v)) \oplus \gamma] = 1,$$

$$BF[(h_i(v))][1 - H(h_{k+1}(h_i(v)) \oplus \gamma] = 0,$$

where $v = Args||CreateTime||Nounce$ is retrieved from the proposed transaction *Tx*, and $i \in [1, k]$, $k$ represents the number of hash mapping functions.
3) *Check*: To confirm the authenticity of the transaction $Tx'$ received from participants, check $v' = Args'||CreateTime'||Nounce'$ within $Tx'$ as:

$$BF\left[(h_i(v'))\right]\left[H(h_{k+1}(h_i(v')) \oplus \gamma\right] = 1.$$

If $v'$ of $k$ selected hash functions for *BF* is found to be 1, it indicates that $Tx'$ is linked to the generation of product *P*.

**Algorithm 1** Details of *CTrace*

**Input:** $BF$ in the world state, transaction $Tx$ related to product $P$, and transaction $Tx'$ to be verified
**Output:** Whether $Tx'$ is associated with product $P$
*Setup*:
Generate two $m$-bit sets as $BF$;
$H_{key} = \{h_1, h_2, \cdots, h_{k+1}\}$;
$H(x) = \mathbb{H}\%2$;
Update $BF$ in the world state for product $P$;
*Insert*:
Propose a transaction $Tx$ related to $P$;
$v = Args||CreateTime||Nounce$;
**for** $i = 1; i \leq k; i++$ **do**
  $BF\left[(h_i(v))\right]\left[H(h_{k+1}(h_i(v)) \oplus \gamma\right] = 1$;
  $BF\left[(h_i(v))\right]\left[1 - H(h_{k+1}(h_i(v)) \oplus \gamma\right] = 0$;
**end for**
Update $BF$ in the world state for product $P$;
*Check*:
Retrieve $Args'$ and $CreateTime'$ from $Tx'$;
$v = Args'||CreateTime'||Nounce'$;
**for** $i = 1; i \leq k; i++$ **do**
  **if** $BF\left[(h_i(v^{'}))\right]\left[H(h_{k+1}(h_i(v^{'})) \oplus \gamma\right] == 0$ **then**
    return 0;
  **end if**
**end for**
return 1;

For instance, in the content generation based on diffusion models, the core workflow can be decomposed into two sequential phases: Forward Diffusion and Reverse Denoising, which collaboratively accomplish content generation through a multi-round iterative process. During the Forward Diffusion phase, key parameters are immortalized on the blockchain, encompassing noise schedule $\beta_1, \beta_2, \ldots, \beta_T$, the set of noisy images $x_1, x_2, \ldots, x_T$, and noise samples $\epsilon$ that are introduced. In the subsequent Reverse Denoising phase, blockchain records the set of denoised images $x_T, x_{T-1}, \ldots, x_0$ and the set of evolving parameters for the predicted noise $\epsilon_\theta(x_t, t)$. Furthermore, considering the request transaction $Tx_{req}$ and multiple content exchange transactions $Tx_{exchange}$, we assume that the number of associated transactions per generated product is constrained by $n \leq 20$. To address the correctness verification requirements, when the false positive rate threshold is set to 0.1%, probability model derivation yields the optimal parameter combination $(k = 10, m = 288)$ satisfying:

$$P_{fp} \approx \left(1 - e^{-kn/m}\right)^k \leq 0.001,$$

This configuration guarantees that the system upholds the necessary probabilistic assurances, striking a balance between computational efficiency and the reliability of verification.

It is worth noting that the query results of *CTrace* are exclusively used for preliminary screening of transaction relevance. Isolated matching results caused by false positives do not possess the validity of legal evidence. In subsequent verification procedures, it is necessary to further retrieve complete on-chain transaction records for verification. The forged transactions corresponding to false positive results cannot provide legitimate digital signatures and thus lack legal effect.

#### Lifecycle management of AIGC product

In the AIGC-Chain, we store all data generated throughout the AIGC creation process in the form of transactions on the blockchain.

We incorporate *User Sign* into the world state of AIGC products, establishing a direct connection between each product and the personal information of producers. This enhances the traceability of the AIGC products, allowing for the efficient monitoring of user identity information, including that of the producer. As a result, the measure effectively prevents the misuse of AIGC models and significantly reduces the production of fraudulent or low-quality AIGC content. To prevent duplicate transactions of product copyright, we add *Owner Sign* in the world state of AIGC products, which ensures that each copyright transaction is permanently recorded on the blockchain. *Consumer*s can review the transaction history of the product and modify *Owner Sign* upon completion of the deal. The transaction process is contingent upon the validation of the ownership signature; only when the signature is successfully verified can the transaction proceed.

To elucidate the full lifecycle recording process of AIGC-Chain with greater clarity, we illustrate the process with the generation of AIGC products via diffusion models as an example. Figure 5 graphically depicts the transition of the world state of the product, with the modifications since the last transaction highlighted in red.

In *Registration*, CA registers roles participating in AIGC-Chain as lightweight or full nodes. *Content Generation* takes the prompt provided by *User* to the system and assigns unique *Product ID* as *key* for the aigproduct *P*. Instead of storing the actual prompt, it saves the hash of the prompt as *Description Hash* in the world state. After completing the content generation, *Provider* sends the hash value and other related generation information of *P* through *Data Uploading* and returns *P* to *User*. *Consumer* and *User* execute copyright transactions and ownership transfers for AIGC product through *Copyright Trading*. In the event of a copyright dispute, *Auditor*s can invoke *Copyright Management* to retrieve the information related to *P* from AIGC-Chain. Following this, the auditors can proceed to analyze and evaluate the copyright of the product. The detailed introductions of the above five stages are as follows:

(1) *Registration*: Users participating in AIGC-Chain are required to register on the blockchain network, with digital certificates *DC* and public-private key pairs (*pk*, *sk*) being issued by the Certificate Authority (CA) on the chain. Only users who have completed CA authentication are granted access to view and participate in blockchain activities. Within this architecture, *User* and *Consumer* participate as lightweight nodes, predominantly hosted on personal computers and mobile devices. They can engage with blockchain interactions without being responsible for data storage. In contrast, *Provider* and *Auditor*, armed with more robust computing and storage power, serve as full nodes within the network, tasked with the responsibility of synchronizing the entire data on the blockchain.

(2) *Content Generation*: When using an AIGC model to generate product *P*, *User* inputs the prompt *Args* into the model and selects a unique *Product ID* as *key* for *P*. Next, *User* calls the smart contract to initiate a transaction $Tx_{req}$ to update the world state for *P* with *key*. As illustrated in Fig. 5, transaction $Tx_{req}$ updates the world state by updating the *Description Hash*, *User Sign*, *Owner Sign*, *BF*, and *Status*. Since the world state is public to all participants on the consortium chain, we use the hash of the prompt *Description Hash* in $Tx_{req}$ to prevent the prompt from being copied or maliciously used by an attacker.

The specific process of this stage is as follows:

**Algorithm 2** Details of content generation

---

**Input:** *Product ID*, prompt *Args*
**Output:** transaction $Tx_{req}$
*User*:
$\sigma_1 = sig_{sk_{User}}(Args)$;
Send $\sigma_1$ and *Args* to *Provider*;
*Provider*:
Verify $\sigma_1$ with $pk_{User}$;
$\sigma_2 = sig_{sk_{Provider}}(Args)$;
Send $\sigma_2$ back to *User*;
*User*:
$Description\ ID = H(Args)$;
Verify the legality of $\sigma_2$ with $pk_{Provider}$;
$Owner\ Sign = sig_{sk_{User}}(Description\ Hash)$;
Propose transaction $Tx_{req}$ to update the world state with $key = Product\ ID$;
Retrieve $Txid_{req}$ from transaction $Tx_{req}$ and store it locally or within a decentralized storage system.

---

- *User* computes $\sigma_1 = sig_{sk_{User}}(Args)$ and sends $\sigma_1$ along with *Args* to *Provider* for content generation.
- *Provider* verifies $\sigma_1$ with $pk_{User}$ of *User*. The successful verification suggests that the prompt *Args* aligns with the intent of the *User* and has not been altered during transmission.
- *Provider* generates $\sigma_2 = sig_{sk_{Provider}}(Args)$ and sends it back to *User*. This indicates that *Provider* acknowledges the legitimacy of the received *Args*, enabling further content generation operations.
- *User* selects a random hash function *H* and computes $Description\ ID = H(Args)$.
- *User* verifies the legality of $\sigma_2$; a successful validation implies the formation of consensus between *User* and *Provider*.
- *User* generates signature *Owner Sign* with *Description Hash*.
- A transaction $Tx_{req}$ is proposed by *User* to update the world state with $key = Product\ ID$, where *Status* is used to indicate the current state of *P*, which is *prepared*.
- *User* retrieves the transaction ID $Txid_{req}$ from transaction $Tx_{req}$ and may choose to store it either locally or within a decentralized storage system such as IPFS,[4] which facilitates the simplification of the copyright management process.

[4] https://www.ipfs.tech.

**Algorithm 3** Details of data uploading

**Input:** $Product\ ID$, prompt $Args$, and ID of the chosen model $Model\ ID$
**Output:** product $P$, transaction $Tx_{gen}$, and transaction $Tx_{upload}$
$\underline{Provider}$:
Generate product $P$ based on the prompt $Args$;
$\sigma_3 = sig_{sk_{Provider}}(P)$;
Send $\sigma_3$ and generated product $P$ back to $User$;
$\underline{User}$:
Verify $\sigma_3$ with $pk_{Provider}$;
$\sigma_4 = sig_{sk_{User}}(P)$;
Send $\sigma_4$ back to $Provider$;
$\underline{Provider}$:
Verify the legality of $\sigma_4$ with $pk_{User}$;
$Model\ Sign = sig_{sk_{Provider}}(steps||seed||Created\ Date)$
Propose transaction $Tx_{gen}$ to update the world state with $key = Product\ ID$;
Retrieve $Txid_{gen}$ from transaction $Tx_{gen}$ and store it locally or within a decentralized storage system.
$\underline{User}$:
$Product\ Hash = H(P)$;
Propose transaction $Tx_{upload}$ to update the world state with $key = Product\ ID$;
Retrieve $Txid_{upload}$ from transaction $Tx_{upload}$ and store it locally or within a decentralized storage system.

(3) *Data uploading*: *Provider* generates product $P$ with the AIGC model based on the prompt *Args* provided by *User*. Subsequently, *Provider* transmits $P$ to *User* and initiates a transaction $Tx_{gen}$ to update additional information related to the generation process in the world state. Upon receiving product $P$, *User* initiates transaction $Tx_{upload}$ to add product $P$ into the world state. Figure 5 illustrates the alteration in the world state following the execution of transactions $Tx_{gen}$ and $Tx_{upload}$. In this process, the use of *Product Hash* in place of the original product $P$ is aimed at preventing potential copyright infringement or misuse, while also ensuring the security of the data in AIGC-Chain.

- *Provider* generates product $P$ based on the prompt *Args* and then computes signature $\sigma_3 = sig_{sk_{Provider}}(P)$ for it.
- *Provider* sends both $\sigma_3$ and the generated product $P$ back to *User* for verification.
- *User* authenticates the validity of $\sigma_3$; only upon successful verification will *User* accept product $P$ and proceed to sign it as $\sigma_4 = sig_{sk_{User}}(P)$, which will be sent back to *Provider*.
- *Provider* verifies the legality of $\sigma_4$; a successful validation implies the formation of consensus between *User* and *Provider*.
- *Provider* generates signature *Model Sign* with data related to the generation of the AIGC product, including *steps*, *seed*, and *Created Date*, which are data used and produced during the AIGC generation process.
- *Provider* proposes a transaction $Tx_{gen}$ to update generation data related to $P$ with $key = Product\ ID$, where *Status* has shifted from *prepared* to *generating*.
- *User* computes $Product\ Hash = H(P)$ for product $P$.
- A transaction $Tx_{upload}$ is proposed by *User* to update the world state, where *Status* has transitioned from *generating* to *generated*.
- *User* and *Provider* retrieve transaction ID $Txid_{gen}$ and $Txid_{upload}$ from the blockchain respectively, and can choose to store them either locally or within a decentralized storage system such as IPFS.

(4) *Copyright trading*: When *Consumer* plans to purchase the copyright of product *P*, it initially retrieves product details from the world state and verifies the validity of product *P*. Following successful validation, *Consumer* makes a payment to the owner and initiates a transaction $Tx_{exchange}$ on the blockchain to update *Owner Sign* and *BF* within the global state. The specific details are as follows:

- *Consumer* solicits *Product ID* of aiming product *P* and the public key $pk_{owner}$ from the owner.
- *Consumer* retrieves *Owner Sign* and *Status* from the world state with *Product ID*, validating the legality of *Owner Sign* with $pk_{owner}$. Simultaneously, it is necessary to ensure that *Status* is not *prepared* or *generating*.
- A transaction $Tx_{exchange}$ is proposed by *Consumer* to update the world state.

(5) *Copyright management*: For a copyright dispute, *Auditor* can retrieve *Product ID* of the disputed product *P* by consulting the relevant world state on the AIGC-Chain. *Provider* and *Producer* are required to provide proof to *Auditor* to establish their connection with the disputed product. *Auditor* will then assess the ownership of the copyright based on the proof gathered.

- *User* submits the proof $\varphi$ to *Auditor*, where $\varphi = (pk_{User}, Txid_{req}, Txid_{upload}, prompt)$.
- *Provider* delivers the proof $\phi$ to *Auditor*, where $\phi = (pk_{Provider}, Txid_{gen}, metadata)$, *metadata* refers to some auxiliary information within the generation process of AIGC, including execution logs, intermediate results, etc.
- *Auditor* utilizes $pk_{User}$ and $pk_{Provider}$ to verify *User Sign* and *Model Sign* retrieved from the world state of the AIGC-Chain for *Product ID*, to assess the correctness of the identities associated with the product-related roles.
- *Auditor* employs *CTrace.Check* to verify the correctness of $Txid_{req}$, $Txid_{upload}$, and $Txid_{gen}$. An output of 1 indicates that the transaction is indeed related to the generation process of the AIGC product.
- *Auditor* conducts an in-depth manual review of the originality, compliance, ownership, and value of the AIGC product based on the information provided by *User* and *Provider*, as well as the product information stored on AIGC-Chain.

## Theoretically evaluation

This section conducts a theoretical evaluation of the performance of the designed system about the design goals outlined in "Models" section.

**Theorem 1** *The copyright-related data for AIGC products retrieved from the blockchain can be utilized to authenticate the copyright ownership of an AIGC product.*

***Proof*** Blockchain is a linked series of blocks, with data dispersed across a network of nodes in a distributed manner. This distribution necessitates that any alteration to the data must be reflected across the majority of nodes, a process that is prohibitively challenging in practice. Additionally, the integration of a new block into the blockchain requires consensus from all nodes, which involves recomputing the necessary tasks and securing the validation and agreement of participants in the network. This process is both resource-intensive and infeasible for malicious purposes. Consequently, the data recorded on the blockchain is reliable and inviolable. This inherent security and immutability make the blockchain an enduring and trustworthy platform for recording the full lifecycle of AIGC products, providing reliable, immutable information for effective copyright management of AIGC products. □

**Theorem 2** *It is impossible for anyone to attempt to alter the copyright-related data of an AIGC product by fabricating transactions on the blockchain.*

***Proof*** *CTrace* establishes a clear association between AIGC product and the transaction ID provided by *User* and *Provider*. Since forged *Txid′* cannot pass *CTrace*.

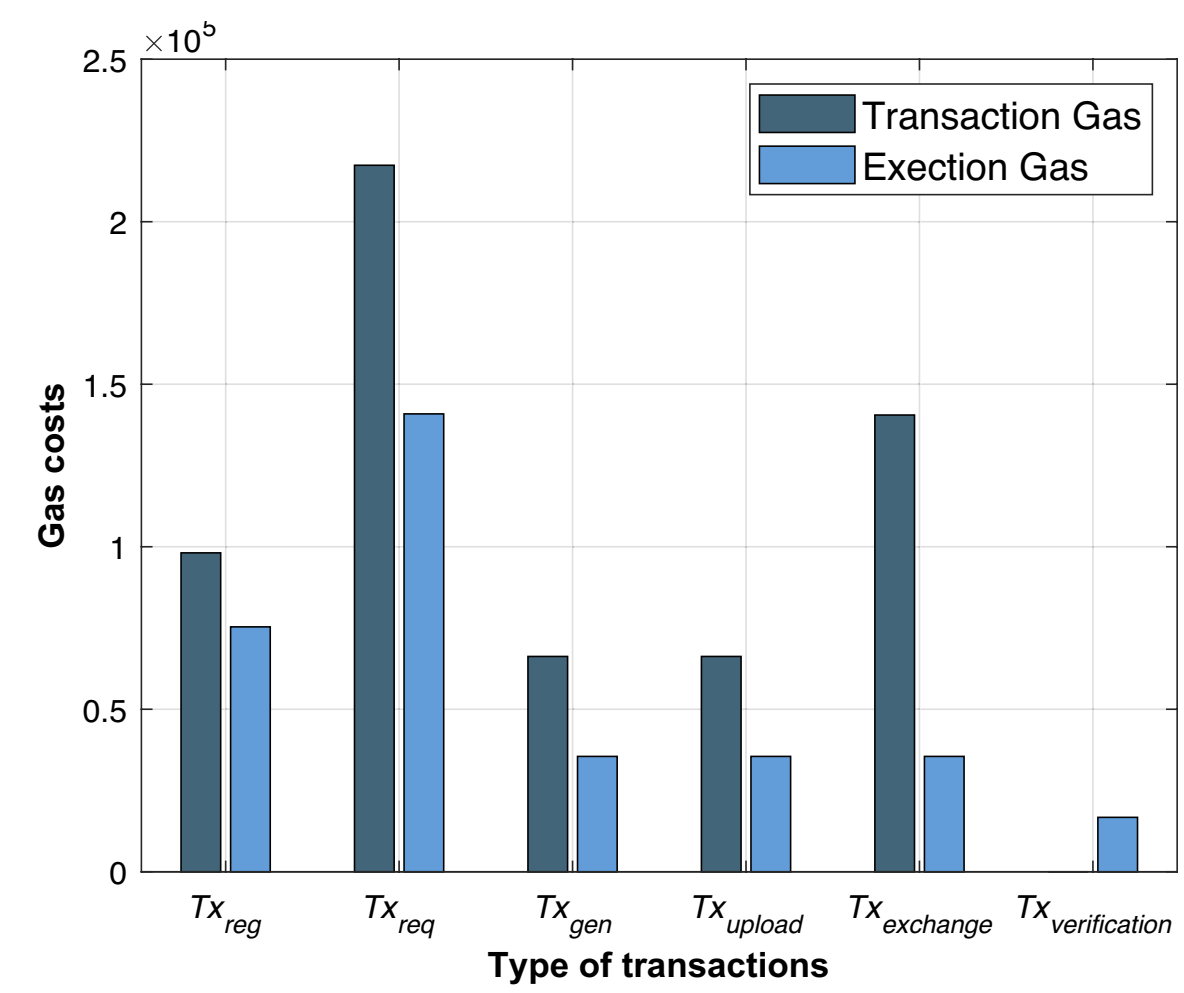

**Fig. 6** Performance analysis of AIGC-chain

check(), preventing them from being mapped within the blockchain, thereby confirming their irrelevance to the AIGC product. This mechanism effectively hinders attackers from posting fraudulent transactions on the blockchain to interfere with the accurate determination of copyright ownership for AIGC products. □

**Theorem 3** *Auditors are equipped to make fair copyright ownership decisions by the decentralized and unchangeable properties of the blockchain.*

***Proof*** *Auditor*s responsible for copyright management are composed of copyright certification institutions, courts, and other authoritative bodies, which also serve as nodes within the AIGC-Chain network. They rely on dependable, immutable data from the blockchain. The decentralized nature of the blockchain necessitates consensus among all nodes for audit conclusions, significantly reducing the risk of malicious auditor actions and ensuring the fairness and security of copyright management. Furthermore, as participants on the blockchain, *Auditors* face the potential reduction in their weight following errors, which could negatively impact their standing and interests. This creates a motivation for auditors to uphold impartiality and strive for precision, ensuring that their judgments are accurate, equitable, and reliable. □

**Theorem 4** *Parameters* $m = 288$ *and* $k = 10$ *are feasible and secure for CTrace under a false positive probability of* 0.1%, *ensuring efficient and reliable copyright verification for AIGC products.*

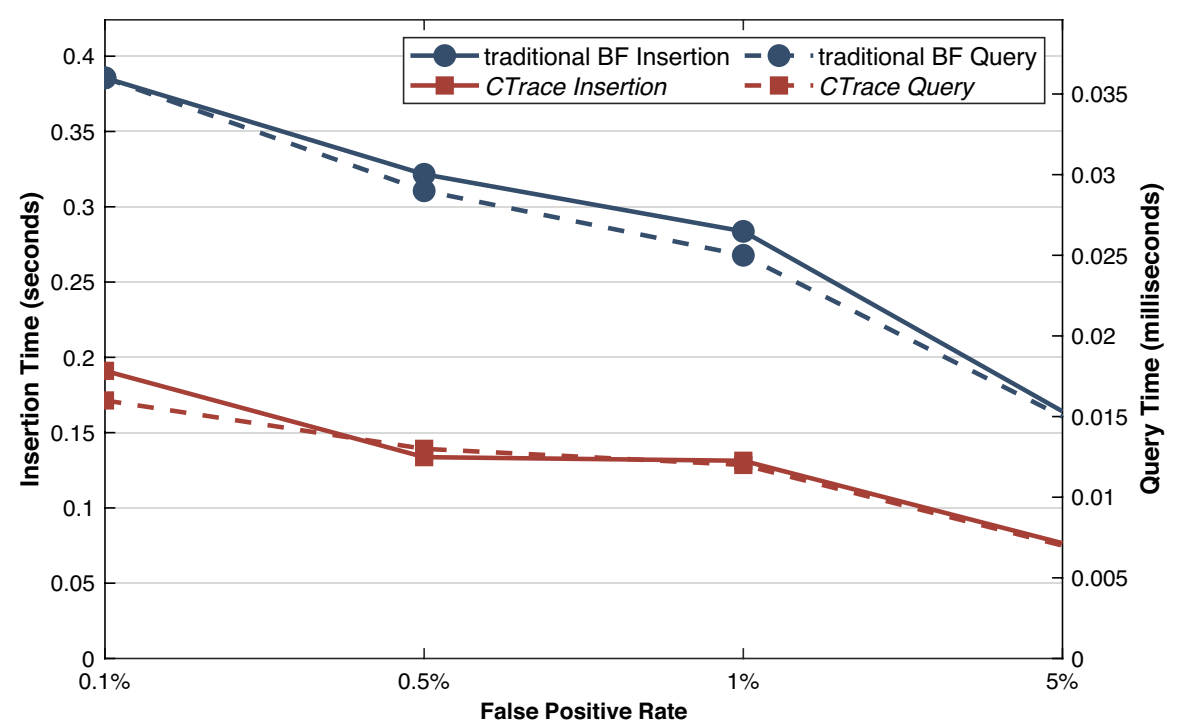

**Fig. 7** Execution efficiency of *CTrace*

***Proof*** The security and feasibility of *CTrace* depend on its ability to balance false positive probability $P_{fp}$, the size of the array $m$, and the number of hash functions $k$. For a given $n$ (number of elements inserted), the false positive rate is bounded by:

$$P_{fp} = \left(1 - e^{-\frac{kn}{m}}\right)^k$$

To achieve $P_{fp} \leq 0.1\%$, we derive the relationship between $m$, $k$, and $n$. The optimal number of hash functions $k$ minimizes $P_{fp}$ for fixed $m$ and $n$:

$$k_{\text{opt}} = \frac{m}{n} \ln 2$$

For $n \leq 20$, substituting $m = 288$ and $k = 10$:

$$\begin{aligned} P_{fp} &= \left(1 - e^{-\frac{10\times 20}{288}}\right)^{10} \\ &= \left(1 - e^{-0.694}\right)^{10} \\ &\approx (1 - 0.500)^{10} = 0.500^{10} \approx 9.76 \times 10^{-4} \end{aligned}$$

This yields $P_{fp} \approx 0.0976\%$ , which meets the requirement $P_{fp} \leq 0.1\%$. The parameters are therefore feasible.

For $P_{fp} = 0.0976\%$, an attacker attempting to spoof $10^4$ AIGC products would trigger $\sim 9.76$ false positives on average. This low rate makes systematic exploitation statistically impractical.

The bit density $\rho$ of *CTrace* is:

$$\rho = 1 - e^{-\frac{kn}{m}} = 1 - e^{-\frac{10\times 20}{288}} \approx 0.500$$

With $\rho \approx 50\%$, the filter maintains stable performance for $n \leq 20$, avoiding accelerated false positive growth.

Using $k = 10$ hash functions introduces $O(k)$ operations for insertion/query. Modern cryptographic hash functions (e.g., SHA-3) execute these operations in $\mu$ s-scale, ensuring real-time compatibility with blockchain transactions.

Therefore, the parameters $m = 288$ and $k = 10$ satisfy $P_{fp} \leq 0.1\%$ for $n \leq 20$, maintaining security and feasibility. □

**Table 2** Comparison of time complexity

| Method | Normal | Fast query | LightChain Hassanzadeh-Nazarabadi et al. (2021) | ADIP Rasheed et al. (2022) | FlexIM Li et al. (2025) | *CTrace* |
|---|---|---|---|---|---|---|
| Time complex | $O(m)$ | $O(n)$ | $O(\log N)$ | $O(\lambda)$ | $O(\log n)$ | $O(1)$ |

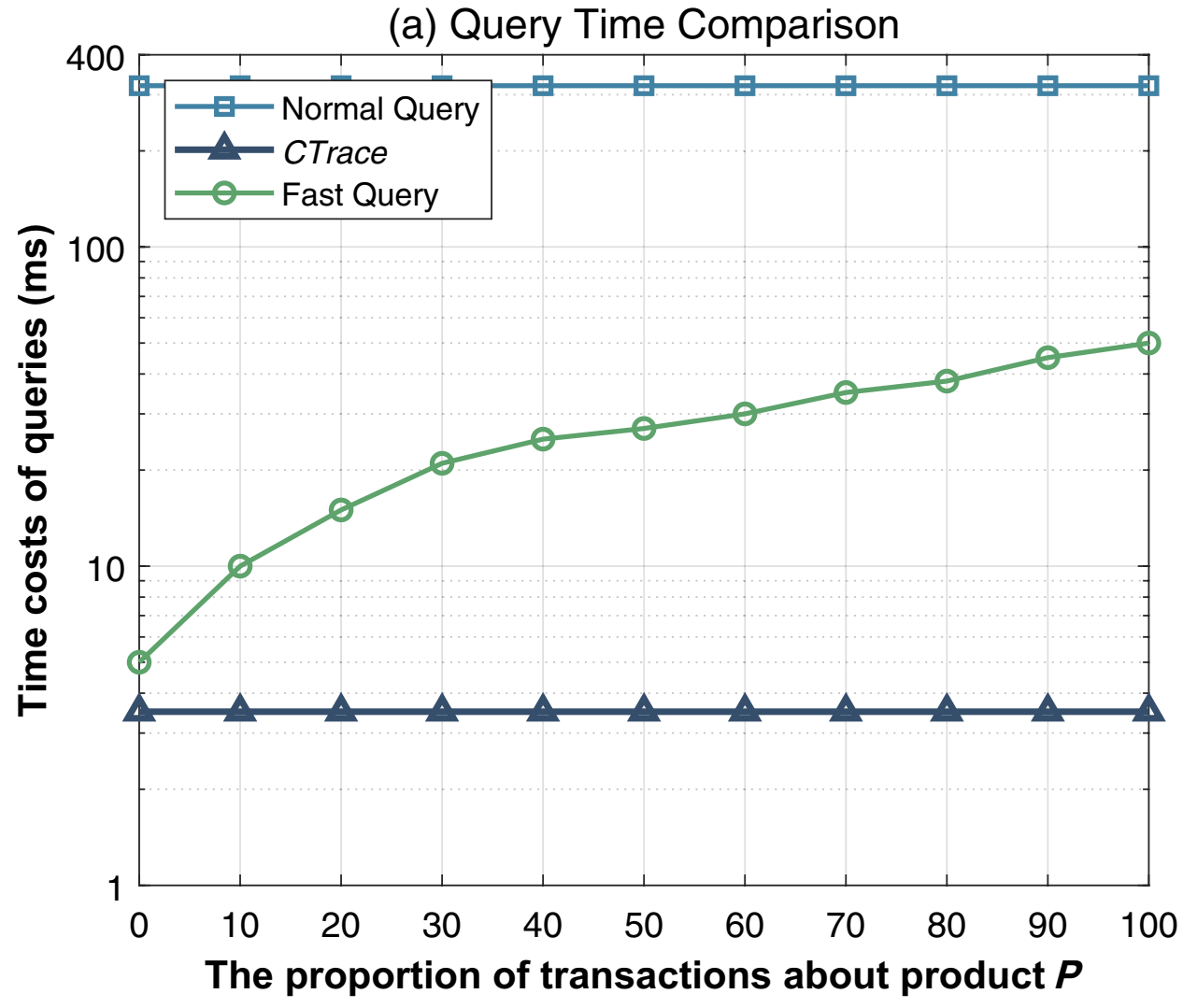

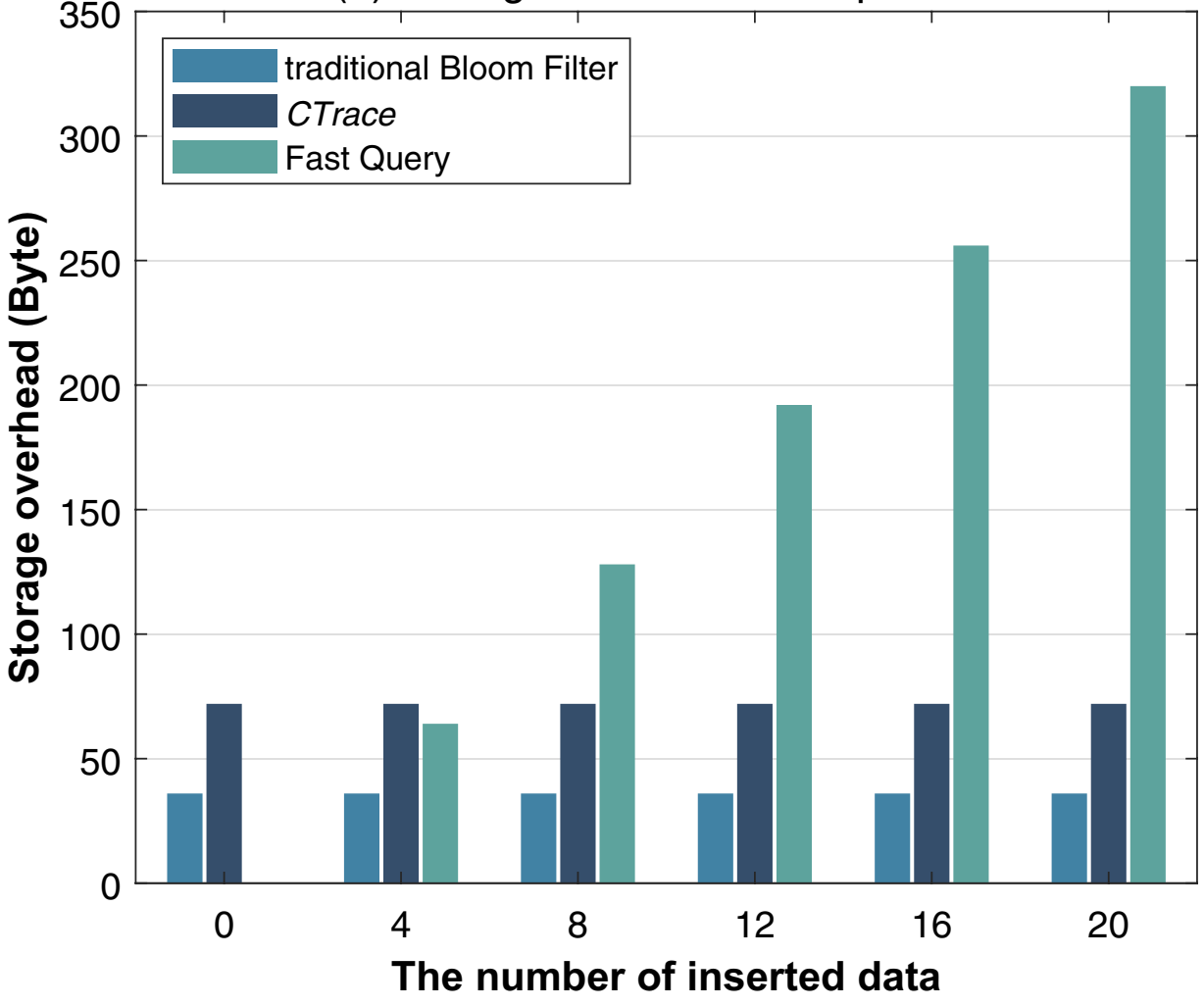

**Fig. 8** Performance analysis of *CTrace*

**Theorem 5** *An attacker cannot feasibly manipulate CTrace to corrupt copyright verification proof.*

***Proof*** The security of *CTrace* relies on the immutability of its bit array and the irreversibility of hash functions. To force a false positive, an attacker must either:

- Reverse-engineer hash outputs: Cryptographic hashes (e.g., SHA-256) are preimage-resistant, making it infeasible to find an input $x'$ that collides with a legitimate copyright entry $x$.
- Overwhelm the filter: Saturating the filter requires inserting $n'$ elements such that $P_{fp}$ exceeds 0.1%. For $m = 288$, achieving $P_{fp} > 0.1\%$ requires:

$$n' > \frac{m\ln(1-p^{1/k})}{-k} = \frac{288 \times \ln(1-(10^{-3})^{1/10})}{-10} \approx \frac{288 \times (-0.693)}{-10} \approx 20.0$$

To trigger a false positive by forging transactions, an attacker would need to insert more than 20 correlated transactions into the blockchain. However, such anomalous insertion attempts are promptly detected and blocked by the consensus mechanism of blockchain, rendering false positives nearly impossible to induce in practical scenarios. □

## Performance evaluation

In this section, we conduct a comprehensive performance analysis and experimental verification of AIGC-Chain, with a specific focus on the performance evaluation of *CTrace*.

### Implementation overview

AIGC-Chain demonstrates cross-platform compatibility with mainstream blockchain platforms. To objectively evaluate system performance metrics, we constructed a blockchain testnet with PoA. The blockchain core is developed in Go (Golang v1.19), while smart contracts are implemented as Solidity (v0.8.0) bytecode executed. Our experimental network architecture comprises 8 validator nodes and one bootnode, with each node instance containerized in isolated Docker environments (v20.10.18) to ensure resource partitioning. All network components operate on KVM-based virtual machines configured with 4GB DDR4 memory and Ubuntu 22.04 LTS. Besides, the size of the Bloom filter used in *CTrace* was set to $m$=288, and the false positive rate was set at 0.1%.

### Blockchain evaluation

The cost efficiency of AIGC-Chain is evaluated based on the amount of gas required for each contract function. We analyze computational costs for various procedures within AIGC-Chain, meticulously documenting transaction costs and execution costs. Transaction gas represents the actual amount of gas expended, whereas execution gas refers to the gas required specifically for the function execution, excluding the overhead associated with the transaction itself. The gas fee consumed is specifically referred to as transaction gas, and it is only charged for functions that alter the state of the blockchain. Read-only functions, on the other hand, do not consume gas and are not subject to latency.

Figure 6 provides a comparative analysis of the transaction gas costs associated with various procedures within the AIGC-Chain. It is observed that the gas cost for $Tx_{req}$ reaches a maximum of 217,370 gas units, which is notably higher than the costs incurred by the subsequent procedures. This increased expenditure is due to the resource-intensive write operations required for the generation of an AIGC product on the blockchain, including the registration of metadata, the generation of cryptographic hashes, and updates to the immutable ledger. Conversely, $Tx_{verification}$ incurs no transaction costs, as it is conducted off-chain through decentralized verification protocols and does not require any on-chain data writing. The metric for execution gas cost reflects the computational resource usage during the execution of smart contracts, including both algorithmic processing and on-chain data retrieval. As depicted in Fig. 6, the AIGC-Chain exhibits a varied distribution of execution gas costs across its 5 main workflow stages. $Tx_{req}$ accounts for the highest execution costs, attributable to the intricate contract logic that involves multi-signature authentication and decentralized storage mechanisms.

### Evaluation of CTrace

This research conducts a comparative evaluation of *CTrace* and Fast Query (Chen et al. 2023), with a focus on their respective optimization strategies for data retrieval. *CTrace* facilitates rapid transaction verification by leveraging a bloom filter within the world state in a systematic manner. In contrast, Fast Query optimizes search efficiency by using a single linked list to index only the copyright transactions related to product *P*. Traditional methods, however, necessitate exhaustive scans of the data on the blockchain for query resolution. The time complexity for executing queries using these methods is detailed in Table 2, where $m$ denotes the total number of transactions on the blockchain, $n$ signifies the number of transactions relevant to product $P$, $N$ denotes the total number of nodes in the blockchain network, and $\lambda$ represents the security parameter. This comparative analysis quantitatively demonstrates the superiority of *CTrace* in data verification.

In the performance evaluation phase, we conducted rigorous comparative experiments between traditional Bloom filters and *CTrace*. Each experimental group involved standardized sequences of 10,000 insert and query operations separately, with each experiment repeated 30 times to eliminate systemic errors through averaging. As quantitatively demonstrated in Fig. 7, *CTrace* exhibits measurable throughput degradation compared to conventional Bloom filters. This performance differential primarily stems from its intrinsic privacy-preserving mechanisms. While this design decision incurs additional computational costs, it successfully prevents the avenues for retroactively inferring the original data through statistical analysis techniques. Nevertheless, this performance trade-off is deemed reasonable and acceptable within the domain of privacy-enhanced data structures.

Furthermore, we compared the performance of *CTrace* with traditional query methods and Fast Query, which relies on a single linked list, focusing on query time and storage overhead. Figure 8a illustrates the query time costs of these three methods under varying data volumes. Traditional methods require invoking smart contracts to access transaction data on the blockchain, resulting in a consistent but significant time cost. Fast Query, with its off-chain single-linked list storage, experiences increasing query times with the expansion of data. In contrast, *CTrace*, by only accessing the bloom filter stored in the world state and performing a single hash mapping operation, maintains a constant query time regardless of storage size. Figure 8b depicts the storage costs under different data volumes, highlighting that storage requirements of *CTrace* remain constant and do not increase with data accumulation, in contrast to Fast Query. As the amount of data increases, *CTrace* will significantly reduce storage overhead compared to Fast Query, demonstrating superior space efficiency.

## Conclusion

This paper confronts the copyright challenges associated with AIGC and introduces a trustworthy management system for AIGC copyright. It meticulously records intermediate data throughout the full lifecycle of AIGC and anchors it within a decentralized blockchain, establishing a robust multi-party auditing framework for the fair and reliable management of AIGC Copyright. The paper further develops a copyright tracing approach utilizing bloom filters, named *CTrace*, which markedly improves the efficiency and accuracy of blockchain transaction query and verification, providing a solid foundation for resolving copyright disputes. These innovations robustly safeguard legitimate rights and offer a reliable approach for the collection and verification of proof in AIGC copyright disputes, promoting the sustainable growth and vibrancy of AIGC. As future directions, we plan to divide the copyright management of AIGC into three distinct phases: model training, requirement formulation, and content generation. Our goal is to explore the copyright issues that may arise in each phase and propose effective strategies to address them.

**Acknowledgements**
Not applicable.

**Author contributions**
Conceptualization and investigation, Fengshu Li. Formal analysis and scheme design, Jiajia Jiang. Secure proof and performance analysis, Xiangli Xiao. Writing-original draft preparation, Jiajia Jiang and Yushu Zhang. Writing-review and editing, Moting Su. Content guidance, Aiqun Wu. All authors have read and agreed to the published version of the manuscript.

**Funding**
This work was supported in part by the Postgraduate Research & Practice Innovation Program of Jiangsu Province under Grant KYCX25_0634, in part by the National Natural Science Foundation of China under Grant 72561011, in part by the Ganpo Talent Program of Jiangxi Province under Grant gpyc20240012, in part by the Outstanding Youth Fund Program of Jiangxi Province under Grant 20252BAC220008, in part by the Jiangxi Key Research and Development Program under Grant 20261BCE310050, in part by Youth Foundation of Philosophy and Social Science Research Planning of Anhui Province under Grant AHSKQ2021D12, and in part by Research Project of Humanities and Social Sciences of the Ministry of Education under Grant 25YJC630055.

**Data availability**
Not applicable.

## Declarations

**Competing interests**
The authors declare no competing interests.

## Publisher's Note

Springer Nature remains neutral with regard to jurisdictional claims in published maps and institutional affiliations.